\documentclass{article}

\usepackage[T2A]{fontenc}
\usepackage[cp866]{inputenc}
\usepackage[english,russian]{babel}
\usepackage{amsmath,amsfonts,amssymb,amscd,mathrsfs}

\usepackage[all]{xy}

\DeclareMathOperator{\PGL}{PGL}

\newcommand{\mathset}[1]{{\left\{#1\right\}}} 
\newcommand{\absolute}[1]{\lvert#1\rvert}

\newcommand{\Fin}{{\rm fin}}
\newcommand{\dist}{{\rm dist}}

\newtheorem{thm}{Theorem}[section]

\newtheorem{Quest}[thm]{Questions}

\begin{document}

\centerline{\large\bf Mumford dendrograms and discrete $p$-adic symmetries}
\medskip

\centerline{\bf P.E.\ Bradley}
\medskip

\centerline{\it Universit\"at Karlsruhe, Germany}
\centerline{\it bradley@ifib.uni-karlsruhe.de}

\selectlanguage{english}
\medskip
\centerline{\today}

\begin{abstract}
In this article, we present an effective encoding of den\-dro\-grams by 
embedding
them into the Bruhat-Tits trees associated to $p$-adic number fields.
As an application, we show how strings over a finite alphabet can be encoded in
cyclotomic extensions of $\mathbb{Q}_p$
and discuss $p$-adic DNA encoding. The application leads to fast $p$-adic
agglomera\-ti\-ve 
hierarchic algorithms similar to the ones recently used e.g.\ by
A. Khrennikov and others. From the viewpoint of $p$-adic geometry,  to encode 
a dendrogram $X$ in a $p$-adic field $K$ means to fix a set $S$ of
$K$-rational punctures on the $p$-adic projective line $\mathbb{P}^1$.
To $\mathbb{P}^1\setminus S$ is associated in a natural way a subtree inside
the Bruhat-Tits tree which recovers $X$, a method first used by F.\ Kato in
1999 in the classification of discrete subgroups of $\textrm{PGL}_2(K)$.

Next, we show how the $p$-adic moduli space $\mathfrak{M}_{0,n}$ of
$\mathbb{P}^1$ with $n$ punctures can be applied to the study of time series
of dendrograms and those symmetries arising from hyperbolic actions on
$\mathbb{P}^1$. In this way, we can associate to certain classes of dynamical
systems a  Mum\-ford curve, i.e.\ a $p$-adic algebraic curve with totally
degenerate re\-duc\-tion modulo $p$.

Finally, we indicate some of our results in the study of general discrete
actions on $\mathbb{P}^1$, and their relation to $p$-adic Hurwitz spaces.
\end{abstract}

\section{Introduction}

Mumford curves arise as the generalisation of the so-called 
{\em Tate uni\-formisation} of $p$-adic elliptic curves
\cite[\S 6]{Tate1974}. The latter has a combinatorial
description as a $\mathbb{Z}$-action on the real line
``connecting'' the points $0$ and $\infty$ over $\mathbb{Q}_p$.
The crucial idea by Mumford \cite{Mumford-DC}
was to view the real line as a geodesic
line inside the Bruhat-Tits tree
$\mathscr{T}_{\mathbb{Q}_p}$ for $\PGL_2(\mathbb{Q}_p)$
and to consider a discrete action of a subgroup 
$G$ generated by
$g$ hyperbolic fractional linear transformations
acting regularly on a subdomain $\Omega$ of the $p$-adic
Riemann sphere $\mathbb{P}^1$.  It turns
out that the orbit space
 $X=\Omega/G$ is a complete algebraic curve of genus $g$,
and that not all $p$-adic algebraic curves admit such a uniformisation.
A curve of the form $X=\Omega/G$ as above is called a {\em Mumford curve}
or a {\em $p$-adic Riemann surface}. 

Here, we are concerned with the application of $p$-adic
geometry in the analysis of hierarchical data.
From a geometric
viewpoint, the  tree $\mathscr{T}_{\mathbb{Q}_p}$ 
represents the hierarchical organisation
of all $p$-adic numbers, including $\infty$. Namely, a $p$-adic number
can in a natural way be viewed as an infinite path inside the tree 
$\mathscr{T}$ starting from some vertex $v$. 
Two paths starting from $v$ correspond to two $p$-adic numbers
having their first terms coincide in their $p$-adic expansions.
 The more terms they have in common, the
closer they are $p$-adically. Hence, 
for some  given  $p$-adic
numbers, their geodesic paths in $\mathscr{T}_{\mathbb{Q}_p}$
will yield a subtree which hierarchically
represents their proximities.
This motivates the usefulness of the Bruhat-Tits tree
for hierarchical data analysis by finding a way of
 encoding data as $p$-adic numbers. Unfortunately, there
is no natural way of doing this for arbitrary data
other than strings over an alphabet. 

Time series of hierarchical data naturally yield the consideration
of  families of sets of $p$-adic numbers. The corresponding
geometric construct is a {\em moduli space} of such families.
Here, they come in the form of $M_{0,n}$, the $p$-adic
{\em moduli space of $n$-pointed genus zero curves}.
Classically, these and their variants in higher genus
play an important role in string theory, and we expect this
also to be the case in $p$-adic string theory.
However, for data mining, a time series is simply a sequence of points
in $M_{0,n}$, and it would certainly be interesting to
be able to interpolate and have a curve inside the moduli space
in order to say something about the evolution of the time series,
or the probability of a certain behaviour in time.


\section{Dendrograms}

Dendrograms are a certain way of depicting trees arising in the
hier\-ar\-chi\-cal classification of data. 
Their intention is usually to describe
hierarchies found within some given dataset. However,  it is often the
result of imposing hierarchies onto the data, depending on the
choice of a  metric. A lot of work by Fionn Murtagh aims
to find ultrametricity in data in order to reveal underlying hierarchy,
e.g.\
\cite{Murtagh-JoC2004,Murtagh-EurPhysJB2005}.
The reason is precisely the tree-like structure of any ultrametric distance.
From a $p$-adic viewpoint, the following procedure seems natural:

\begin{enumerate}
\item Encode dataset $X=\mathset{x_1,\dots,x_n}$ by $p$-adic numbers $Y$.
\item Construct the dendrogram for $X$ from the code $Y$.
\end{enumerate}  

The dendrogram for $X$ is  uniquely determinded by $Y$ and
can be computed quite fast. Hence, the true problem is
to find a suitable encoding by $p$-adic numbers. This is in general
a very difficult task, as one is likely to need the dendrogram {\em a priori}.
However, for strings of letters from a given alphabet, we will
show how  $p$-adic encodings can be effected in Section \ref{pstrings}.

A more precise definition of a dendrogram is that of a metrised tree with
finitely many ends, all of which are labelled.

In what follows, we assume that 
to each dataset $X$, there is given a dendrogram $D$ 
which is supposed to reveal the hierarchical structure within $X$.

\section{$p$-adic dendrograms}

\begin{figure}[h]
$$
\xymatrix@R=2pt@C=2pt{
&&&&&&&&&&\infty\ar@{-}[d]&&&&
\\
0\ar@{-}[r]&*\txt{}\ar@{-}[ddddddd]&&&&&&*\txt{}\ar@{-}[dd]&&&*\txt{}\ar@{-}[rrr]^1\ar@{-}[lll]_0&&&*\txt{}\ar@{-}[d]&
\\
1\ar@{-}[r]&*\txt{}&&&&&&&&&&&*\txt{}\ar@{-}[dddddd]&*\txt{}\ar@{-}[l]_1\ar@{-}[r]^0&*\txt{}\ar@{-}[dddddd]
\\
2\ar@{-}[r]&*\txt{}&&&&*\txt{}\ar@{-}[ddd]&&*\txt{}\ar@{-}[ll]_0\ar@{-}[rr]^1&&*\txt{}\ar@{-}[d]&&&&&
\\
3\ar@{-}[r]&*\txt{}&&&&&&&*\txt{}\ar@{-}[d]&*\txt{}\ar@{-}[l]_0\ar@{-}[r]^1&*\txt{}\ar@{-}[dddd]&&&&
\\
4\ar@{-}[r]&*\txt{}&&&&&&*\txt{}\ar@{-}[ddd]&*\txt{}\ar@{-}[l]_0\ar@{-}[r]^1&*\txt{}\ar@{-}[ddd]&&&&&
\\
5\ar@{-}[r]&*\txt{}&&&*\txt{}\ar@{-}[d]&*\txt{}\ar@{-}[l]_0\ar@{-}[r]^1&*\txt{}\ar@{-}[dd]&&&&&&&&
\\
6\ar@{-}[r]&*\txt{}&&*\txt{}\ar@{-}[d]&*\txt{}\ar@{-}[l]_0\ar@{-}[r]^1&*\txt{}\ar@{-}[d]&&&&&&&&&
\\
&&&x_1&&x_2&x_3&x_4&&x_5&x_6&&x_7&&x_8
\\
}
$$
\caption{A $2$-adic dendrogram.} \label{dendro2ad}
\end{figure}
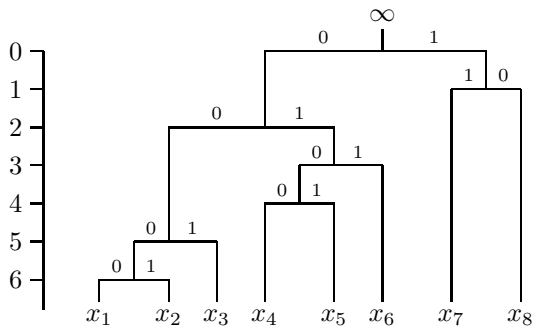

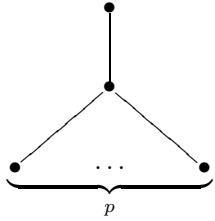
\begin{figure}[h]
$$
\underbrace{
\xymatrix{
&*\txt{$\bullet$}&\\
&*\txt{$\bullet$}\ar@{-}[u]\ar@{-}[dl]\ar@{-}[dr]&\\
*\txt{$\bullet$}&\dots&*\txt{$\bullet$}\\
}
}_{p}
$$
\caption{Local structure of $\mathscr{T}_{\mathbb{Q}_p}$.}
\label{plocaldisc}
\end{figure}

Consider the dendrogram $D$ as depicted in Figure \ref{dendro2ad}.
If one goes down from $\infty$ along a path in $D$ to some datum $x=x_i$,
and picks up the labels $0$ or $1$ along the way, then one gets
a $2$-adic encoding 
$$
x=\sum\limits_{\text{\; level $\nu$}}a_\nu 2^\nu \in\mathbb{Q}_2,
$$
where coefficient $a_\nu$ is the number picked up at level $\nu$.
Here, this yields the numbers
$$
\begin{array}{llll}
x_1=0, & x_2=2^{6},
& x_3=2^5,& x_4=2^2,
\\
x_5=2^2+2^4,& x_6=2^2+2^3,& x_7=2^0+2^1,& x_8=1
\end{array}
$$
Notice that the labels are arranged here in such a way that the code
will be $x_1=0$  at the very left,
and $x_n=1$ at the very right (and $\infty$ on top) of the dendrogram.
Of course, the procedure yields
 just  finite $2$-adic expansions of  rational numbers.
But in this way, the whole dendrogram $D$ gets embedded into
an infinite tree: the {\em Bruhat-Tits tree} $\mathscr{T}_{\mathbb{Q}_2}$
for the group $\PGL_2(\mathbb{Q}_2)$. 
For a general prime number $p$,
the tree $\mathscr{T}_{\mathbb{Q}_p}$ is a locally finite $p+1$-regular
tree. The latter 
means that from each vertex there are precisely $p+1$ emanating
edges. The reason is that the vertices can be interpreted as $p$-adic
discs, and the edges are given by maximal non-trivial inclusion
of discs.  It is known that each disc has precisely $p$ maximally
smaller subdiscs and lies inside precisely one minimal bigger disc.
Hence, each vertex has precisely $p$ children vertices and one
parent vertex. This is illustrated in Figure \ref{plocaldisc}.

The number $p$ of children vertices comes from
the isomorphism $\mathbb{Z}_p/p\mathbb{Z}_p\cong\mathbb{F}_p$,
saying that the residue field of $\mathbb{Q}_p$ is the finite
field $\mathbb{F}_p$. 
Hence, each downward edge can be labelled
by any representative  for $\mathbb{F}_p$ in 
the ring $\mathbb{Z}_p$ of $p$-adic integers.
Quite common is e.g.\ the set of labels $\mathset{0,\dots,p-1}$.
 
Moving downwards from some vertex $v$ on will end in some $p$-adic
number  $x\in\mathbb{Q}_p$ as the intersection of a
decreasing sequence of discs corresponding to the vertices on the
infinite path, and picking up labels as before, yields
a $p$-adic expansion of $x$ as a Laurent series in $p$.
Hence, all of $\mathbb{Q}_p$ can be considered as 
lying at the boundary of $\mathscr{T}_{\mathbb{Q}_p}$.
However, there is one more boundary point of $\mathscr{T}_{\mathbb{Q}_p}$
outside $\mathbb{Q}_p$: taking the path going upwards from each vertex
will lead to the point $\infty$. Hence, we have found
$$
\partial\mathscr{T}_{\mathbb{Q}_p}=\mathbb{Q}_p\cup\mathset{\infty}
=\mathbb{P}^1(\mathbb{Q}_p),
$$
where the latter space $\mathbb{P}^1$ is the $p$-adic
{\em projective line}. 

We have seen even more that the local picture of the Bruhat-Tits tree allows
a local interpretation as another projective line:
namely, there is a bijection
$$
\mathset{\text{\rm edges emanating from vertex $v$}}\cong
\mathbb{F}_p\cup\mathset{\infty}=\mathbb{P}^1(\mathbb{F}_p),
$$
with the projective line  defined  this time over the residue field
$\mathbb{F}_p$.

Let now be given a finite set $X=\mathset{x_0,\dots,x_n}$ of
$p$-adic numbers.  Then by taking inside the Bruhat-Tits tree $\mathscr{T}_{\mathbb{Q}_p}$
all geodesic paths between the points of $X$, one obtains a subtree
$\mathscr{T}(X)$ which we call a {\em $p$-adic dendrogram}. We give credit
to Fumiharu Kato who used this construction already in 1999 in the classification of $p$-adic discrete projective linear groups 
(cf.\ also \cite[\S 5.2]{Kato-JAG}).

Observe that the $2$-adic encoding of a dendrogram described above
yields a $2$-adic dendrogram for the $2$-adically coded data
plus the extra ``datum'' $\infty$. This extra point at infinity
allows to determine the root of a dendrogram, from which all paths
to genuine data are oriented downwards, i.e.\ passing through children
vertices. In Figure \ref{dendro2ad}, the root corresponds to the
unit disc, because on the one hand
 the data code contains $0,1\in\mathbb{Q}_2$,
and the three numbers $0,1,\infty$ uniquely determine the unit disc.
And on the other hand, all data are encoded by numbers within the unit disc.

\section{Non-binary data}

In the previous section, we have seen how to $p$-adically
encode data having a {\em binary} dendrogram,
and we have defined  $p$-adic dendrograms which are not
necessarily binary. Hence, a natural way of encoding data
whose dendrogram $D$
 is not binary would be by increasing the prime number $p$
to
the  size of at least the maximal  number of children vertices in $D$.

However, there is an alternative way of doing this without changing the
prime $p$. Namely, consider
 any finite field extension $K$ of $\mathbb{Q}_p$.
It is well-known that the $p$-adic norm extends uniquely
 to a norm $\absolute{\ }_K$,
and that $K$ is complete for this norm.
Again, the unit disc is the ring 
$O_K=\mathset{x\in K\mid \absolute{x}_K\le 1}$,
and the next smaller disc containing $0$ is
$\pi O_K$, where $\pi$ is a so-called {\em uniformiser}
and plays the role of the prime $p$ in $K$.
It holds true that the residue field
$$
\kappa:=O_K/\pi O_K\cong \mathbb{F}_q
$$
is a finite field extension of $\mathbb{F}_p$ 
with $q=p^f$ elements for some natural number $f\ge 1$.

In general, it holds true that 
\begin{align}
f=\dim_{\mathbb{F}_p}(\kappa)\le\dim_{\mathbb{Q}_p}(K)=:n, \label{resdegree}
\end{align}
where the dimensions are of that of vector spaces over the scalar fields
$\mathbb{Q}_p$ and $\mathbb{F}_p$, respectively.
The result is that there are more discs defined over $K$ than over 
$\mathbb{Q}_p$. More precisely, the number of ``children'' disks 
has increased to $q=p^f$, and there is a new Bruhat-Tits tree
$\mathscr{T}_K$ which is again infinite, but this time $q+1$-regular.
The analogue holds true: 
\begin{align*}
\partial\mathscr{T}_K&\cong \mathbb{P}^1(K)
\\
\mathset{\text{edges emanating from vertex $v$}}&\cong \mathbb{P}^1(\kappa).
\end{align*}
In fact, there is an embeding of trees
$$
\mathscr{T}_{\mathbb{Q}_p}\to\mathscr{T}_K
$$
given in general  by subdividing edges and increasing the 
number of edges emanating from a vertex.
Note that that the subdivision of edges comes from
a relation between the uniformisers:
$$
\absolute{\pi}_K^e=\absolute{p}_p,
$$
which causes the length of an edge in $\mathscr{T}_K$ to be 
an $e$-th fraction of an edge length in $\mathscr{T}_{\mathbb{Q}_p}$.
The number $e$ is called the {\em ramification index} of the field extension
$K/\mathbb{Q}_p$. By adopting the labelling method from above,
we obtain the general encoding
$$
x=\sum\limits_{\nu=-m}^\infty a_\nu\pi^\nu,
$$
where $a_\nu$ is taken from a system $\mathcal{R}$ 
of representatives in $K$ for the residue field $\kappa$.
This is nothing but the $\pi$-adic expansion of elements from $K$.

In the case that (\ref{resdegree})
is an equality, the field extension $K/\mathbb{Q}_p$
is called {\em unramified}. By the well known formula
$$
n=e\cdot f,
$$
this is equivalent to $e=1$.
In this case, the prime $p$ can be taken as the uniformiser
of $K$, and we obtain again $p$-adic expansions---only with more
choice of coefficients.
A special case is given by a so-called {\em cyclotomic} extension
$K=\mathbb{Q}_p(\zeta)$ obtained by adjoining to $\mathbb{Q}_p$
the powers of a primitive 
$(p^f-1)$-th root $\zeta$ of unity. This case is known to be unramified,
and we can take as coefficients 
\begin{align}
\mathcal{R}_f:=\mathset{0,\zeta,\dots,\zeta^{p^f-2}}
\label{teichmueller}
\end{align}
for the $p$-adic expansion of elements from $K$.
Note, that for $f=1$, this yields a  set of coefficients
different from the usual choice $\mathset{0,\dots,p-1}$.

The proofs for most of the statements in this section can be found
in \cite[Ch.\ 5]{Gouvea}. 

\section{Strings over an alphabet} \label{pstrings}

Let $\mathcal{A}$ be a finite alphabet.
We will show how to realise $p$-adic encodings of strings over $\mathcal{A}$.

\medskip
First, denote by $S(\mathcal{A})$ the set of all strings with 
letters from $\mathcal{A}$. The subset of finite strings will be denoted by 
$S_\Fin(\mathcal{A})$.
Now, for $f$ sufficiently large, any injective map
$$
\mathcal{A}\to\mathcal{R}_f,
$$
with $\mathcal{R}_f$ defined as in (\ref{teichmueller}),
induces an encoding of $S(\mathcal{A})$ in $O_K$, where $K$ is the
cyclotomic field $\mathbb{Q}(\zeta)$ with $\zeta$ a primitive $(p^f-1)$-th
root of unity. Clearly, the finite strings are then in bijection with
the set of polynomials $\mathcal{R}_f[p]$
in the prime $p$ whose coefficients are from $\mathcal{R}_f$. 
We even have more \cite[Thm.\ 3.1]{Brad-TCJ}:

\begin{thm}
There exists a cyclotomic field  $K=\mathbb{Q}_p(\zeta)$
with $\zeta$ as above, and a closed isometric embedding
$\phi\colon S(\mathcal{A})\to O_K$ such that
$\phi(S_\Fin(\mathcal{A}))\subseteq\mathcal{R}_f[p]$, and is dense in 
$\phi(S(\mathcal{A}))$.
\end{thm}

Here, the metric on $S(\mathcal{A})$ is given by the {\em Baire distance}
$$
\delta_p(x,y):=
\inf\mathset{p^{-n}\mid \text{first $n$ letters of $x$ and $y$ coincide}}
$$
Note that the image $\phi(S(\mathcal{A}))$ is 
a disc with holes coming from the complement of $\phi(\mathcal{A})$
in $\mathcal{R}_f$. More precisely, if $x\in O_K$
is represented as
$$
x=\sum\limits_{\nu\in\mathbb{N}}a_\nu p^\nu,\; a_\nu\in\mathcal{R}_f,
$$
the holes are given as the union of open discs
\begin{align*}
\mathset{a_0\notin\phi(\mathcal{A})}
\cup
\mathset{a_1\notin\phi(\mathcal{A})}
\cup
\mathset{a_2\notin\phi(\mathcal{A})}
\cup\dots
\end{align*}
Note further, that although 
there are only finitely many encodings $\phi\colon \mathcal{A}\to\mathcal{R}_f$,
there are infinitely many $p$-adic encodings by changing the system 
$\mathcal{R}\subseteq O_K$
of representatives for the residue field $\kappa$.

\section{$p$-adic clustering} \label{pcluster}

If data $X$ are encoded $p$-adically, it is a very simple and fast task to
retrieve the uniquely determined hierarchical structure of $D$
given by the tree $\mathscr{T}(X)$.
Any clustering algorithm using the $p$-adic metric will never need to
change the metric when measuring distances between disjoint clusters
$C_1$ and $C_2$, because 
of the fact 
$$
\dist_p(C_1,C_2)=\absolute{x-y}_p
$$
for any $x\in C_1$, $y\in C_2$. Essentially, the fact that one seeks
a subtree of a tree makes things more simple and faster
than in the archimedean situation. In \cite[\S 3]{Brad-TCJ},
an explicit form of a $p$-adic hierarchic classification algorithm has been
discussed. 
Benois-Pineau et al.\ have applied such an algorithm in image segmentation
\cite{BPXK2001}.

\section{DNA}

As an example for what has been said in the previous sections,
we discuss
 $p$-adic encoding of DNA.
Here, the alphabet is given as $\mathcal{A}=\mathset{A,G,C,T}$,
where 
\begin{align*}
A&= \text{Adenine}
&G&= \text{Guanine}\\
C&=\text{Cytosine}
&T&=\text{Thymine}
\end{align*}

Dragovich and Dragovich 
\cite{DD-genomics} choose a
$5$-adic encoding in the field $\mathbb{Q}_5$
$$
\phi_{\rm DD}\colon\mathcal{A}\to\mathcal{R}=\mathset{0,1,2,3,4}
$$
with $\phi_{\rm DD}(\mathcal{A})=\mathset{1,2,3,4}$. This allows for taking
$0$ as a ``blank'' in order to separate words made out of $\mathcal{A}$.
So, in fact, they use the extended alphabet
$\mathcal{A}\cup\mathset{\text{``blank''}}$ and encode it with a bijection to
$\mathcal{R}$ taking the ``blank''  to $0$.

Khrennikov and Kozyrev \cite{XK-genomics} use a bijection
$$
\mathcal{A}\to\mathbb{F}_2^2
$$
as their encoding. As an $\mathbb{F}_2$-vector space,
$\mathbb{F}_2^2$ is isomorphic to the additive group of the
finite field
$\mathbb{F}_{2^2}$ with four elements. 
This field, in turn, is the residue field
of the cyclotomic field $K=\mathbb{Q}_2(\zeta)$
with $\zeta$ a primitive third root of unity.
Because of the correspondence
$$
1\leftrightarrow \begin{pmatrix}1\\0\end{pmatrix},\quad
\zeta\leftrightarrow\begin{pmatrix}0\\1\end{pmatrix},\quad
1+\zeta\leftrightarrow\begin{pmatrix}1\\1\end{pmatrix},
$$
 their encoding can be interpreted as choosing
$\mathcal{R}_{\rm XK}=\mathset{0,1,\zeta,1+\zeta}$ 
and a bijection
$$
\phi_{\rm XK}\colon\mathcal{A}\to\mathcal{R}_{\rm XK},
$$
which gives a $2$-adic encoding. However, there is no ``blank''
in this case.

By the previous sections, we see that there are a lot more pos\-si\-bi\-li\-ties,
even for $2$-adic encodings. For a version without ``blank'',
a bijection with 
$$
\mathcal{R}_2=\mathset{0,1,\zeta,\zeta^2}
$$
could be used. And for a version with ``blank'',
an injection into 
$$
\mathcal{R}_3=\mathset{0,1,\xi,\dots,\xi^6}
$$
could be interesting,
where $\xi$ is a seventh root of unity\footnote{Notice that $7=2^3-1$.}.

The following questions come up naturally:
\begin{Quest}
1. Are there among the possible $2$-adic encodings
$$
\mathcal{A}\to\mathcal{R}_3
$$
some more preferred than others from the point of view
of genomics?

2.
Which are the best choices for systems $\mathcal{R}\subseteq O_K$
of representa\-ti\-ves for the residue field $\kappa=\mathbb{F}_{2^3}$
from a genomic point of view (possibly including ``blank'')?
\end{Quest}

Of course, there is the question, whether cyclotomic 
or unramified $p$-adic fields
are sufficiently suited for genomics. 

\section{Time series}

Assume that we are given some time  dependent $p$-adically
encoded data, i.e.\ a set of $p$-adic numbers
$$
S_t=\mathset{s_0(t),\dots,s_n(t)}
$$
at some instances
of time $t=0,1,\dots,N$. We assume that $\infty\in S_t$
and that there are no ``collisions'' at any time,
if we may use the language of ``particles'' moving inside some
``space''.
This corresponds to the
$p$-adic projective line with $n+1$ points removed:
$$
X_t:=\mathbb{P}^1\setminus S_t,
$$
which is the usual way of denoting an {\em $n+1$-pointed genus zero curve}.
If we normalise for each $t$
 via some fractional linear map
$$
z\mapsto\frac{az+b}{cz+d}
$$
the punctures $S_t$  to contain $0,1,\infty$, we have 
in $X_t$ a standard re\-pre\-sen\-ta\-ti\-ve of a point
$x_t$ inside the {\em moduli space $M_{0,n+1}$ of $n+1$-pointed
genus zero curves} defined over $\mathbb{Q}_p$.
In the language of moduli spaces, the time series $S_t$
corresponds to a family $X_t$ of punctured curves
which in turn comes from a map
$$
\mathset{0,1,\dots,N}\to M_{0,n+1}.
$$
Collisions can also be treated in this way:
simply replace $M_{0,n+1}$ by a suitable com\-pac\-ti\-fi\-ca\-tion
$\bar{M}_{0,n+1}$ in which the boundary corresponds to all possible
ways of colliding particles.

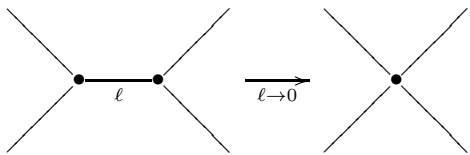
\begin{figure}[h]
$$
\xymatrix{
&&&&&&\\
&*\txt{$\bullet$}\ar@{-}[lu]\ar@{-}[ld]\ar@{-}[r]_\ell
&*\txt{$\bullet$}\ar@{-}[ur]\ar@{-}[dr]&
\ar[r]_{\ell\to 0}&&
*\txt{$\bullet$}\ar@{-}[ur]\ar@{-}[dr]\ar@{-}[ul]\ar@{-}[dl]
\\
&&&&&&
}
$$
\caption{Edge contraction.} \label{edgecontract}
\end{figure}

There is now an infinite-to-one map
$$
\Pi\colon M_{0,n+1}\to \mathcal{D}_n,\; \mathbb{P}^1\setminus S\mapsto T(S)
$$
into the space $\mathcal{D}_n$ of all dendrograms for $n$ data and $\infty$.
The fibre of a point  $x\in\mathcal{D}_n$ corresponds to the
infinitely many possible $p$-adic encodings of the dendrogram associated 
to $x$. Hence, these correspond to the
{\em sections} $f\colon\mathcal{D}_n\to M_{0,n}$, i.e.\ maps
satisfying 
$$\Pi\circ f={\rm id}_{\mathcal{D}_n}.
$$

The  space $\mathcal{D}_n$ is a polyhedral complex of dimension
$$
\dim\mathcal{D}_n=\dim M_{0,n+1}=(n+1)-3,
$$
where the subtraction of $3$ comes from the normalisation after which
$3$ points are fixed.
The maximal cells $\mathcal{D}_n$ are all of the dimension of the
moduli space and consist of the binary dendrograms.
A cell in $\mathcal{D}_n$ is characterised by  the fact
that the abstract trees corresponding to its elements are all isomorphic,
whereas the edge lengths vary. Passing to a neighbouring cell
amounts to contracting an edge as illustrated in Figure \ref{edgecontract}.

\section{Genus 1 time series}

\begin{figure}[h]
$$
\xymatrix@C=10pt{
&&\infty&\\
&*\txt{}\ar@{-}[rr]\ar@{-}[dd]&*\txt{$\bullet$}\ar@{-}[u]&*\txt{}\ar@{-}[ddd]\\
&&&\\
*\txt{}\ar@{-}[rr]\ar@{-}[d]&*\txt{$\bullet$}&*\txt{}\ar@{-}[d]&
\\
0&&x&1
}
\quad
\xymatrix@C=10pt{
&&\infty&\\
&*\txt{}\ar@{-}[rr]\ar@{-}[d]&*\txt{$\bullet$}\ar@{-}[u]&*\txt{}\ar@{-}[ddd]
\\
*\txt{}\ar@{-}[rr]\ar@{-}[dd]&*\txt{$\bullet$}&*\txt{}\ar@{-}[dd]&
\\
&&&
\\
0&&x&1
}
\quad
\xymatrix@C=10pt@R=23pt{
&\infty&&\\
*\txt{}\ar@{-}[rr]\ar@{-}[ddd]&*\txt{$\bullet$}\ar@{-}[u]&*\txt{}\ar@{-}[d]
\\
&*\txt{}\ar@{-}[rr]\ar@{-}[dd]&*\txt{$\bullet$}&*\txt{}\ar@{-}[dd]
\\
&&&&\text{etc.}
\\
0&x&&1
}
$$
\caption{A sequence of dendrograms.} \label{dendroseries}
\end{figure}
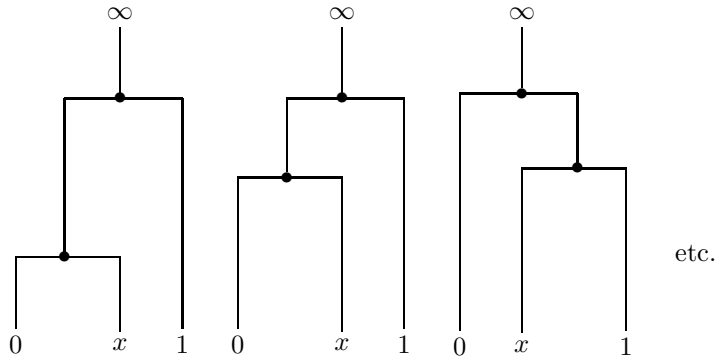

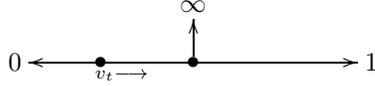
\begin{figure}[h]
$$
\xymatrix@R=10pt{
&&\infty&&\\
0\ar[rrrr]&*\txt{$\bullet$}\ar[l]^{\hspace*{17mm} v_t\longrightarrow}&*\txt{$\bullet$}\ar[u]&&1\\
}
$$
\caption{Vertex jump.} \label{vertexjump}
\end{figure}

Consider the sequence of dendrograms as given in Figure \ref{dendroseries}.
We can view this as a vertex $v_t$ determined by $x$ at time $t$
``jumping'' along the geodesic line between $0$ and $1$ with respect
to the fixed vertex determined by $\infty$, as in Figure \ref{vertexjump}.

If the vertex jumps at a constant rate, i.e.\ the distance at each time
is the same, then we can model this by a translation along that geodesic line,
or $p$-adically via a M\"obius transformation
$$
\gamma\colon z\mapsto\frac{-1}{(1-c)z-1}\in\PGL_2(K),
$$
where $\absolute{c}<1$. This  corresponds for $v_t$ to a jump 
of distance $-\log_p\absolute{c}$ to the right.
This is the case of $\gamma$ being hyperbolic.

By a change of coordinates taking $(0,1,\infty)$ to $(0,\infty, 1)$
we trans\-form everything said above to a hyperbolic action 
of the cyclic group $\langle\gamma\rangle$ on
the geodesic line between $0$ and $\infty$, or $p$-adically: on
$K^\times=\mathbb{P}^1\setminus\mathset{0,\infty}$.
Hence,  $\gamma$ has now the form
$$
\gamma\colon z\mapsto c\cdot z,
$$
and we obtain the commutative diagram

\begin{align*}
\xymatrix@C=17pt@R=10pt{
&K^\times\ar[rrrr]^\gamma\ar@{~>}[dd]&&&&K^\times/\langle\gamma\rangle=E
\ar@{~>}[dd]\\
&&&&&\\
&&&&&
}
\\[-2.5mm]
&\hspace*{-7cm}
\xymatrix{
0\ar[rr]&&\infty\ar[ll]
}
\;\;
\xymatrix{
\ar[rr]_\gamma&&
}
\quad
\xymatrix{
*\txt{$\bullet$}\ar@{-}@(dr,ur)
}
\end{align*}
in which $E$ is a so-called {\em Tate elliptic curve}. It is a 
$p$-adic curve of genus $1$, and the vertical wiggly arrows
are the so-called {\em reduction} or {\em tropicalisation maps}.

The above example generalises to the case of a discrete action
on the $p$-adic projective line $\mathbb{P}^1$
of a group $G$ of fractional linear transformations inside
$\PGL_2(K)$. If, in this case $\Omega\subseteq\mathbb{P}^1$
is the domain on which $\Gamma$ acts without limit or fixed points,
then $C=\Omega/\Gamma$ is known to be a so-called {\em Mumford curve},
a $p$-adic analogon of {\em Riemann surface}.

\section{Identifying $p$-adic Riemann surfaces}

Mumford curves are considered   for $p$-adic
higher genus string ampli\-tu\-des in \cite{CMZ1989}, where the authors
call them {\em $p$-adic Riemann surfaces}. Unlike  in the classical case,
not all  algebraic curves defined over $\mathbb{Q}_p$ are 
$p$-adic Riemann surfaces. However, Chekhov et al.\ conjecture
that the other curves do not contribute
to the $p$-adic (or adelic) string amplitude \cite[Conjecture \S 4.3]{CMZ1989}.
This brings  another physical motivation to 
the general problem of recognising Mumford curves
among algebraic curves. 

\smallskip
Notice that the loop in the commutative diagram of the preceding
section
 is of length $-\log_p\absolute{c}$ and hence
shrinks to zero, if $\absolute{c}$ approach\-es unity. In this case,
the fractional linear transformation $\gamma$ is not hyper\-bolic,
and there is no longer a discrete action of 
$\langle\gamma\rangle$ on the geodesic.

On the side of elliptic curves, this corresponds to the fact
that the family of elliptic curves parametrised by $\gamma$
converges to a $p$-adic elliptic curve which is not a Tate curve.
Such curves do exist, and they can be distinguished by their $j$-invariant. 

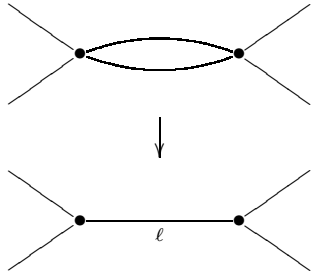
\begin{figure}[h]
$$
\xymatrix@R=15pt{
&&&&\\
&*\txt{$\bullet$}\ar@{-}[ul]\ar@{-}[dl]\ar@/_/@{-}[rr]\ar@/^/@{-}[rr]
&&*\txt{$\bullet$}\ar@{-}[ur]\ar@{-}[dr]&\\
&&\ar[d]&&
\\
&&&&\\
&*\txt{$\bullet$}\ar@{-}[ul]\ar@{-}[dl]\ar@{-}[rr]_\ell
&&*\txt{$\bullet$}\ar@{-}[ur]\ar@{-}[dr]&\\
&&&&
}
$$
\caption{Tate curve covering $\mathbb{P}^1$, graphically.}
\label{Tategraph}
\end{figure}

In fact, let the elliptic curve $E$ be given by an equation over $K$
in Legendre normal form
$$
E\colon y^2=x(x-1)(x-\lambda),
$$
where we may assume that $\absolute{\lambda}=1$
(this implies also $\absolute{\lambda-1}\le 1$).
Then, if $K$ is a sufficiently large finite extension of $\mathbb{Q}_p$,
  it holds true by \cite[Ex.\ 3.8]{Brad-manmat} that
\begin{align*}
\text{$E$ is a Tate curve}
& \Leftrightarrow
\absolute{j(E)}>\absolute{2}_p^4
\\&\Leftrightarrow
\absolute{\lambda-1}<\absolute{2}_p^2
\end{align*}
This result was already known for the case $p>2$, in which
$\absolute{2}_p=1$ \cite[Thm.\ 5]{Tate1974}. 
The last equivalence follows from a well-known
formula relating $\lambda$ and the $j$-invariant.
In order to show that the first and third statements are
equivalent, one can consider the cover $\phi\colon E\to\mathbb{P}^1$
of degree $2$ defined by the Legendre equation:  $\phi$
is simply projection onto the $x$-coordinate.
This induces  a cover of degree $2$ of the tree 
$\mathscr{T}(\mathset{0,1,\infty,\lambda})$ as depicted
in Figure \ref{Tategraph}. The proof then consists of calculating the infimum of $\ell$
for which the upper graph still represents a Tate curve \cite{Brad-manmat}.
For $p=2$, the intuition $\inf(\ell)=0$ fails because of too many fixed
vertices of the elliptic involution on the Bruhat-Tits tree. 

In the case of general Mumford curves
(or $p$-adic Riemann surfaces, if one wishes), one can study the cover
$\Omega\to\Omega/G=C$ in a similar combinatorial way.
Here, the so-called {\em Hurwitz spaces}, which are mo\-du\-li
spaces for covers between curves, come into play. It turns out
that the question whether the upper curve in a cover $f\colon X\to Y$
is a Mumford curve is subtle. Only a restricted type of covers $f$
can in principle  allow $X$ to be a Mumford curve, and 
even then the answer depends on the position of the branch points
of the covering map $f$ \cite{Brad-MZ,Brad-manmat}.

\section{Conclusion}

A $p$-adic encoding of hierarchical data has been discussed from a
geometric point of view. Any dendrogram can in this way be viewed
as a subtree of the Bruhat-Tits tree for $\PGL_2(K)$ defined over a $p$-adic
field $K$ large enough to encapture the maximal number of children vertices
in the dendrogram. This is possible without changing the prime number $p$.
The philosophical result is that cluster analysis becomes
the finding of a suitable $p$-adic encoding of data, because then
the dendrogram is uniquely determined by the ultrametric geo\-metry.
As an example, strings over a finite alphabet have been con\-si\-dered,
where the $p$-adic distance coincides with the Baire distance.
Application to encoding of DNA has been discussed, where the 
general question
is raised which arithmetic conditions on a $2$-adic field $K$
must be imposed from the point of view of genomics.

A consideration of time series of hierarchical data
leads to families of dendrograms or $n$-pointed $p$-adic projective
lines and their moduli spaces as a natural geometric framework.
Higher genus $p$-adic algebraic curves come into the scene, if
a time series can be modelled via a discrete action of
fractional linear transformations on the $p$-adic Riemann sphere.
This and $p$-adic multiloop calculations in string theory
\cite{CMZ1989} motivate the question of how to decide
whether a given algebraic curve of higher genus 
is a $p$-adic Riemann surface. 
 
It is the hope that methods from
$p$-adic string theory and e\-nu\-me\-ra\-ti\-ve geometry will eventually 
find their way into hierarchical data analysis.

\section*{Acknowledgements}

The work presented here is supported by the Deutsche 
For\-schungs\-ge\-mein\-schaft
project BR 3513/1-1. Vladik Avetisov,  Branko Dra\-go\-vich
and Fionn Murtagh are
thanked for helpful questions, remarks and suggestions.
Thanks to the organisers for the wonderful conference.

\end{document}